\documentstyle[twoside]{article}
\catcode`\@=11
\long\def\@makefntext#1{
\protect\noindent \hbox to 3.2pt {\hskip-.9pt
$^{{\eightrm\@thefnmark}}$\hfil}#1\hfill}       

\def\@makefnmark{\hbox to 0pt{$^{\@thefnmark}$\hss}}    
\def\ps@myheadings{\let\@mkboth\@gobbletwo
\def\@oddhead{\hbox{}
\rightmark\hfil\eightrm\thepage}
\def\@oddfoot{}\def\@evenhead{\eightrm\thepage\hfil
\leftmark\hbox{}}\def\@evenfoot{}
\def\sectionmark##1{}\def\subsectionmark##1{}}
\oddsidemargin=\evensidemargin
\newcounter{sectionc}\newcounter{subsectionc}\newcounter{subsubsectionc}
\renewcommand{\section}[1] {\vspace{12pt}\addtocounter{sectionc}{1}
\setcounter{subsectionc}{0}\setcounter{subsubsectionc}{0}\noindent
    {\tenbf\thesectionc. #1}\par\vspace{5pt}}
\renewcommand{\subsection}[1] {\vspace{12pt}\addtocounter{subsectionc}{1}
    \setcounter{subsubsectionc}{0}\noindent
    {\bf\thesectionc.\thesubsectionc. {\kern1pt \bfit #1}}\par\vspace{5pt}}
\renewcommand{\subsubsection}[1] {\vspace{12pt}\addtocounter{subsubsectionc}{1}
    \noindent{\tenrm\thesectionc.\thesubsectionc.\thesubsubsectionc.
    {\kern1pt \tenit #1}}\par\vspace{5pt}}
\newcommand{\nonumsection}[1] {\vspace{12pt}\noindent{\tenbf #1}
    \par\vspace{5pt}}
\newcommand{\textlineskip}{\baselineskip=13pt}

\def\eightcirc{
\begin{picture}(0,0)
\put(4.4,1.8){\circle{6.5}}
\end{picture}}
\def\eightcopyright{\eightcirc\kern2.7pt\hbox{\eightrm c}}

\def\abstracts#1#2#3{{
    \centering{\begin{minipage}{4.5in}\baselineskip=10pt\footnotesize
    \parindent=0pt #1\par
    \parindent=15pt #2\par
    \parindent=15pt #3
    \end{minipage}}\par}}

\newcommand{\bibit}{\nineit}

\renewenvironment{thebibliography}[1]
    {\frenchspacing
     \ninerm\baselineskip=11pt
     \begin{list}{\arabic{enumi}.}
    {\usecounter{enumi}\setlength{\parsep}{0pt}
     \setlength{\leftmargin 12.7pt}{\rightmargin 0pt} 
     \setlength{\itemsep}{0pt} \settowidth
    {\labelwidth}{#1.}\sloppy}}{\end{list}}
\newcounter{itemlistc}
\newcounter{romanlistc}
\newcounter{alphlistc}
\newcounter{arabiclistc}

\def\@citex[#1]#2{\if@filesw\immediate\write\@auxout
    {\string\citation{#2}}\fi
\def\@citea{}\@cite{\@for\@citeb:=#2\do
    {\@citea\def\@citea{,}\@ifundefined
    {b@\@citeb}{{\bf ?}\@warning
    {Citation `\@citeb' on page \thepage \space undefined}}
    {\csname b@\@citeb\endcsname}}}{#1}}
\newif\if@cghi
\def\cite{\@cghitrue\@ifnextchar [{\@tempswatrue
    \@citex}{\@tempswafalse\@citex[]}}
\def\citelow{\@cghifalse\@ifnextchar [{\@tempswatrue
    \@citex}{\@tempswafalse\@citex[]}}
\def\@cite#1#2{{$\null^{#1}$\if@tempswa\typeout
    {IJCGA warning: optional citation argument
    ignored: `#2'} \fi}}

\def\pmb#1{\setbox0=\hbox{#1}
    \kern-.025em\copy0\kern-\wd0
    \kern.05em\copy0\kern-\wd0
    \kern-.025em\raise.0433em\box0}

\def\fnt#1#2{\footnotetext{\kern-.3em
    {$^{\mbox{\scriptsize #1}}$}{#2}}}
\def\fpage#1{\begingroup
\voffset=.3in
\thispagestyle{empty}\begin{table}[b]\centerline{\footnotesize #1}
    \end{table}\endgroup}

\font\tenrm=cmr10
\font\tenit=cmti10
\font\tenbf=cmbx10
\font\bfit=cmbxti10 at 10pt
\font\ninerm=cmr9
\font\nineit=cmti9

\font\eightrm=cmr8

\def\qed{\hbox{${\vcenter{\vbox{            
   \hrule height 0.4pt\hbox{\vrule width 0.4pt height 6pt
   \kern5pt\vrule width 0.4pt}\hrule height 0.4pt}}}$}}
\begin{document}
\normalsize\textlineskip
\thispagestyle{empty}
\setcounter{page}{1}
\vspace*{0.88truein} \fpage{1} \centerline{\bf {\bfit p}-ADIC AND
ADELIC MINISUPERSPACE} \vspace*{0.035truein} \centerline{\bf
QUANTUM COSMOLOGY} \vspace*{0.37truein} \centerline{\footnotesize
GORAN S. DJORDJEVI\'C} \vspace*{0.015truein}
\centerline{\footnotesize\it Sektion Physik, Universit\"at
M\"unchen, Theresienstr. 37, D-80333 M\"unchen, Germany}
\vspace*{0.015truein}\centerline{\footnotesize\it Department of Physics,
University of Ni\v s, P. O. Box 224, 18001 Ni\v s, Yugoslavia}
\vspace*{10pt} \centerline{\footnotesize BRANKO DRAGOVICH}
\vspace*{0.015truein} \centerline{\footnotesize\it Steklov
Mathematical Institute, Gubkin St. 8, GSP-1, 117966, Moscow, Russia}
\vspace*{0.015truein} \centerline{\footnotesize\it Institute of
Physics, P. O. Box 57, 11001 Belgrade, Yugoslavia} \vspace*{10pt}
\centerline{\footnotesize LJUBI\v SA D. NE\v SI\'C}
\vspace*{0.015truein}\centerline{\footnotesize\it Department of Physics,
University of Ni\v s, P. O. Box 224, 18001 Ni\v s, Yugoslavia}
\vspace*{10pt}
\centerline{\footnotesize IGOR V. VOLOVICH} \vspace*{0.015truein}
\centerline{\footnotesize\it Steklov Mathematical Institute,
Gubkin St. 8, GSP-1, 117966, Moscow, Russia} \vspace*{0.225truein}
\abstracts{We consider the formulation and some elaboration of $p$-adic
and adelic quantum cosmology. The adelic generalization of the Hartle-Hawking
proposal does not work in models with matter fields. $p$-adic
and adelic minisuperspace quantum cosmology is well defined as an ordinary
application of $p$-adic and adelic quantum mechanics. It is illustrated by a
few of cosmological models in one, two and three minisuperspace dimensions.
As a result of $p$-adic quantum effects and the adelic
approach, these models exhibit some discreteness of the minisuperspace
and cosmological constant. In particular, discreteness of the de Sitter space
and its cosmological constant is emphasized.}
{}{}
\vspace*{1pt}\textlineskip  
\section{Introduction}  
\vspace*{-0.5pt}
\noindent
The main task of quantum cosmology\cite{wil1} is to describe the evolution of
the universe in a the very early
stage. At this stage, the universe is in a
quantum state, which is described by a wave function. Usually one
takes it that this wave function is complex-valued and depends on some real
parameters. Since quantum cosmology is related to the Planck scale phenomena
it is logical to reconsider its foundations. We will here maintain
the standard point of view that the wave function takes complex values,
but we will treat its arguments in a more complete way. Namely, we will
regard space-time coordinates and matter fields to be adelic, i.e. they
have real as well as $p$-adic properties simultaneously. This approach
is motivated by the following reasons: (i) the field of rational
numbers $Q$, which contains all observational and experimental numerical
data, is a dense subfield not only in the field of real numbers $R$
but also in the fields of $p$-adic numbers $Q_p$ ($p$ is any prime number),
(ii) there is a plausible analysis\cite{vol1} within and over $Q_p$ as well as
that one related to $R$, (iii) general mathematical methods and
fundamental physical laws should be invariant\cite{vol2} under an interchange of
the number fields $R$ and $Q_p$, (iv) there is a quantum
gravity uncertainty\cite{vol2,gar1} $\Delta x$ while measuring distances around the
Planck length $\ell_0$,
\begin{equation}
 \Delta x \geq \ell_0 = \sqrt{\frac{\hbar G}{c^3}} \sim 10^{-33} cm,
\label{1}
\end{equation}
which restricts the priority of archimedean geometry based on real numbers
and gives rise to employment of non-archimedean geometry related to
$p$-adic numbers,\cite{vol2} (v) it seems to be quite reasonable
to extend compact  archimedean geometries by the nonarchimedean ones in the path
integral method, and (vi) adelic quantum mechanics\cite{dra1} applied
to quantum cosmology provides realization of all the above statements.

The successful application of $p$-adic numbers and adeles in modern
theoretical and mathematical physics started in 1987, in the context of
string amplitudes\cite{vol3,fre1} (for a review, see
Refs. 2, 8 and 9). For a systematic research in this field
it was formulated $p$-adic quantum mechanics\cite{vol4,rue1} and
adelic quantum mechanics.\cite{dra1,dra2} They are quantum mechanics
with complex-valued wave functions of $p$-adic and adelic arguments,
respectively. In the unified form, adelic quantum mechanics contains
ordinary and all $p$-adic quantum mechanics.

As there is not an appropriate $p$-adic Schr\"odinger equation,
there is also no $p$-adic generalization of the Wheeler-De Witt
equation. Instead of the differential approach, Feynman's path integral method
will be exploited.

$p$-adic gravity and the wave function of the universe were considered
in the paper\cite{dra3} published in 1991.  An idea of the fluctuating
number fields at the Planck scale was introduced and it was suggested
that we restrict the Hartle-Hawking\cite{har1} proposal to the summation only over
algebraic manifolds. It was shown that the wave function for the de Sitter
minisuperspace model can be treated in the form of an infinite product
of $p$-adic counterparts.

Another approach to quantum cosmology, which takes into account $p$-adic
effects, was proposed in 1995.\cite{dra4} Like in adelic quantum mechanics,
the adelic eigenfunction of the universe is a product of the corresponding
eigenfunctions of real and all $p$-adic cases. $p$-adic wave functions
are defined by $p$-adic generalization of the Hartle-Hawking\cite{har1}
path integral proposal. It was shown that in the framework of this procedure
one obtains an adelic wave function for the de Sitter minisuperspace model.
However, the adelic generalization with the Hartle-Hawking $p$-adic prescription
does not work well when minisuperspace has more than one dimension,
in particular, when matter fields are taken into consideration.
The solution of this problem was found\cite{dra5} by treating minisuperspace
cosmological models as models of adelic quantum mechanics.

In this paper we consider adelic quantum cosmology as an application of
adelic quantum mechanics to the minisuperspace models. It will be illustrated
by one-, two- and three-dimensional minisuperspace models. As a result of
$p$-adic effects and the adelic approach, in these models there is some
discreteness of minisuperspace and cosmological constant. This kind of
discreteness was obtained for
the first time in the context of adelic the de Sitter quantum model.\cite{dra4}

In the next section we give some basic facts on $p$-adic and adelic
mathematics. Section 3 is devoted to a brief review of $p$-adic and
adelic quantum mechanics. $p$-adic and adelic quantum cosmology are
formulated in Sec. 4. Sections 5 and 6 contain some concrete
minisuperspace models. At the end, we give some concluding remarks.

\section{$p$-Adic Numbers and Adeles}
\noindent
We give here a brief survey of some basic properties of $p$-adic numbers
and adeles, which we exploit in this work.

Completion of $Q$ with respect to the standard absolute value
($|\cdot |_\infty$) gives $R$, and an algebraic extension of $R$ makes $C$.
According to the Ostrowski theorem\cite{ost1} any non-trivial norm on
the field of rational numbers $Q$ is equivalent to the absolute value
$|\cdot|_\infty$  or to a $p$-adic norm $|\cdot|_ p$, where $p$ is a prime
number. $p$-adic norm is the non-archimedean (ultrametric) one and for a
rational number $0\ne x\in Q$, where $\ x=p^\nu {m \over n}$,
$\ 0\ne n,\nu, m\in Z$
and $m, n$ are not divisible by $p$,
has a value $|x|_p=p^{- \nu}$. Completion of $Q$ with respect to the $p$-adic
norm for a fixed $p$ leads to the corresponding field of $p$-adic numbers $Q_p$.
Completions of $Q$ with respect to  $|\cdot |_\infty$ and all $|\cdot |_p$
exhaust all possible completions of $Q$.

A $p$-adic number $x\in Q_p$, in the canonical form, is an infinite
expansion
\begin{equation}
x=p^\nu\sum\limits_{i=0}\limits^{+\infty} x_ip^i, \quad
  \nu, x_i \in Z, \quad    0\leq x_i\leq p-1.
\label{2.1}
\end{equation}
The norm of $p$-adic number $x$ in (\ref{2.1}) is $|x|_p=p^{-\nu}$ and
satisfies not
only the triangle inequality, but also the stronger one
\begin{equation}
|x+y|_p\le\max (|x|_p,|y|_p).
\label{2.2}
\end{equation}
Metric on $Q_p$ is defined by $d_p(x,y)=|x-y|_p$. This metric is the
non-archimedean one and the pair ($Q_p,d_p$) presents locally compact,
topologically complete, separable and totally disconnected $p$-adic metric
space.

In the metric space $Q_p$, $p$-adic ball $B_\nu(a)$, with the
centre at the point $a$ and the radius $p^\nu$ is the set
\begin{equation}
\label{2.3}
B_\nu(a)=\{x\in Q_p:\ |x-a|_p\le p^\nu,\ \nu\in Z\}.
\label{2.4}
\end{equation}
The $p$-adic sphere $S_\nu(a)$ with the centre $a$ and the radius
$p^\nu$ is
\begin{equation}
\label{2.5}
S_\nu(a)=\{x\in Q_p:\ |x-a|_p=p^\nu,\ \nu\in Z\}.
\end{equation}
The following holds:
$$
B_\nu(a)=\bigcup_{\nu'\le\nu}S_{\nu'}(a),
$$
$$
S_\nu(a)=B_\nu(a)\backslash B_{\nu-1}(a),\quad
B_\nu(a)\subset B_{\nu'}(a),\quad\nu<\nu',
$$
\begin{equation}
\bigcap_\nu B_\nu(a)=\{ a\},\quad\bigcup_\nu
B_\nu(a)=\bigcup_\nu S_\nu(a)=Q_p.
\label{2.6}
\end{equation}

Elementary $p$-adic functions\cite{sch1} are given by the series of the
same form as in the real case, e.g.
\begin{equation}
\exp x=\sum_{k=0}^\infty \frac{x^k} {k!},
\label{2.7}
\end{equation}
\begin{equation}
\sinh x=\sum_{k=0}^\infty \frac{x^{k+1}} {(2k+1)!},
\qquad
\cosh x=\sum_{k=0}^\infty \frac{x^{2k}} {(2k)!},
\label{2.8}
\end{equation}
\begin{equation}
\tanh x=\sum_{k=2}^\infty \frac{2^k(2^k-1)B_kx^{k-1}} {k!},
\qquad
\coth x= \frac{1} {x}+ \sum_{k=2}^\infty \frac{2^kB_kx^{k-1}} {k!},
\label{2.9}
\end{equation}
where $B_k$ are Bernoulli's numbers. These functions have the same domain of
convergence $G_p=\{ x\in Q_p : |x|_p < |2|_p \} $.
Note the following $p$-adic norms of  the hyperbolic functions:
$|\sinh x|_p=|x|_p$ and $|\cosh x|_p=1$.

Real and $p$-adic numbers are unified in the form of the adeles.\cite{gel1}
An adele is an infinite sequence
\begin{equation}
a=(a_\infty, a_2,...,a_p,...),
\label{2.10}
\end{equation}
where $a_\infty\in Q_\infty$, and $a_p\in Q_p$, with restriction
that $a_p\in Z_p$ $(\ Z_p=\{x\in Q_p:
|x|_p\leq1\})$ for almost all $p$, i.e. for all but
a finite set $S$ of primes $p$.

If we introduce
${\cal A}(S)=Q_\infty\times\prod\limits_{p\in S} Q_p\times\prod
\limits_{p \notin S} Z_p$, then the space of all adeles is
 ${\cal A}=\bigcup\limits_S{\cal A}(S)$, which  is a topological ring.
Namely, ${\cal A}$ is a ring with respect to the componentwise addition and
multiplication. A principal adele is a sequence $(r,r,...,r,...)\in
{\cal A}$, where $r\in Q$. Thus, the ring of principal adeles, which is
a subring of ${\cal A}$, is isomorphic to $Q$.

An important function on ${\cal A}$ is the
additive character $\chi(x), \ x\in {\cal A}$, which is a continuous and
complex-valued function with basic properties:
\begin{equation}
|\chi (x)|_\infty=1,
\quad\chi(x +y)= \chi(x) \chi(y).
\label{2.11}
\end{equation}
This additive character may be presented as
\begin{equation}
\chi(x)=\prod_v\chi_v(x_v)=
\exp(-2\pi ix_\infty)\prod_p \exp(2\pi i\{x_p\}_p),
\label{2.12}
\end{equation}
where $ v=\infty,2,\cdots,p,\cdots$, and $\{x\}_p$ is the fractional
part of the $p$-adic number $x$.

Map $\varphi : {\cal A}\to C$, which has the form
\begin{equation}
\varphi(x) = \varphi_\infty(x_\infty)\prod_{p\in S}\varphi_p(x_p)
\prod_{p\not\in S}\Omega(|x_p|_p),
\label{2.13}
\end{equation}
where $\varphi_\infty(x_\infty)$ is an
infinitely differentiable function on $Q_\infty$ and
falls to zero faster than any power of $| x_\infty |_\infty$ as
$| x_\infty |_\infty \to\infty$,   $\ \ \varphi_p(x_p)$
is a locally constant function with compact
support, and
\begin{equation}
\Omega (|x|_p) = \left\{\begin{array}{ll}
1, & |x|_p \leq 1,\\
0, & |x|_p >1,
\end{array}
\right.
\label{2.14}
\end{equation}
is called an elementary function of ${\cal A}$.
Finite linear combinations of elementary functions (\ref{2.13}) make
the set of the Schwartz-Bruhat functions\cite{gel1} $D({\cal A})$.
The existence of
$\Omega$-function is unavoidable for a construction of any adelic model.
The Fourier transform is
\begin{equation}
\tilde\varphi(\xi) = \int_{\cal A}\varphi(x)\chi(\xi x)dx
\label{2.15}
\end{equation}
and it maps one-to-one $D({\cal A})$ onto $D({\cal A})$. It is worth noting
that $\Omega$-function is a counterpart of the Gaussian
$\exp (-\pi x^2)$ in the real case,
since it is invariant with respect to the Fourier transform.

The integrals\cite{vol1} of the Gauss type over the $p$-adic sphere
$S_\nu = S_\nu (0)$,
$p$-adic ball $B_\nu = B_\nu (0)$ and over any $Q_v$ are:
\begin{equation}
\label{2.16}
\int_{S_\nu}
\chi_p \left( \alpha x^2+\beta x \right)dx =
\left\{\begin{array}{ll}
\lambda_p(\alpha)|2\alpha|_p^{-1/2}
\chi_p \left(
-\frac{\beta^2}{4\alpha}
\right), &
\left|\frac{\beta}{2\alpha}\right|_p=p^\nu,\\
0, & \left|
\frac{\beta}{2\alpha}\right|_p\neq p^\nu,
\end{array}
\right.
\end{equation}
for  $|4\alpha|_p\geq p^{2-2\nu}$,
\begin{equation}
\label{2.17}
\int_{B_\nu}\chi_p(\alpha x^2+\beta x)dx =
\left\{\begin{array}{ll}
p^{\nu}
\Omega(p^{\nu}|\beta|_p), & |\alpha|_pp^{2\nu}\leq1, \\
\frac{\lambda_p(\alpha)}
{|2\alpha|^{1/2}_p}\chi_p\left(-\frac{\beta^2}{4\alpha}\right)
\Omega\left(p^{-\nu}\left|\frac{\beta}{2\alpha}\right|_p\right),
 & |4\alpha|_pp^{2\nu}>1,
\end{array}
\right.
\end{equation}
\begin{equation}
\label{2.18}
\int_{Q_v}\chi_p(\alpha x^2+\beta x)dx= \lambda_v(\alpha)|2\alpha |^{-1/2}_v
\chi_v\left(-\frac{\beta^2}{4\alpha}\right),\quad \alpha \ne 0.
\end{equation}
The  arithmetic functions $\lambda_v(x):\enskip  Q_v\mapsto C$,
where $v= \infty, 2, 3, 5,\cdots$, are defined\cite{vol1} as follows:
$\lambda_v (0) = 1,\quad \lambda_\infty (x) =\frac{1}{\sqrt{2}}(1-i\ sign\ {x})$,
\begin{equation}
\lambda_p(x) = \cases{1,&$\nu = 2k,\ \ \ \ \ \ \ p\not=2,$\cr
\big({x_0\over p}\big),&$\nu = 2k+1,\ \ p\equiv1(\hbox{mod 4}),$\cr
i\big({x_0\over p}\big),&$\nu = 2k+1,\ \ p\equiv3(\hbox{mod 4}),$\cr}
\label{2.19}
\end{equation}
$$
\lambda_2(x) = \cases{{1\over\sqrt2}[1+(-1)^{x_{1}}i],&$\nu = 2k,$\cr
{1\over\sqrt2}(-1)^{x_{1}+x_{2}}[1+(-1)^{x_{1}}i],&$\nu = 2k+1,$\cr}
$$
where $p$-adic $x$ is given by (\ref{2.1}), $k\in Z$, and $\big(\frac{x_0}{p}
\big)$ is the Legendre symbol.
However we will mainly use their properties:
\begin{equation}
|\lambda_v(a)|_\infty=1,\quad
 \lambda_v(ab^2)=\lambda_v(a), \quad
\lambda_v(a)\lambda_v(b)= \lambda_v(a+b)\lambda_v(ab(a+b)).
\label{2.20}
\end{equation}

\section{$p$-Adic and Adelic Quantum Mechanics}
\noindent
In foundations of standard quantum mechanics (over $R$) one
usually starts with a representation of the canonical commutation relation
\begin{equation}
\label{3.1}
[\hat q,\hat k]=i\hbar ,
\end{equation}
where $q$ is a spatial coordinate and $k$ is the
corresponding momentum. It is well known that the procedure of
quantization is not unique.
In formulation of $p$-adic quantum mechanics\cite{vol4,rue1} the multiplication
$\hat q\psi\rightarrow x \psi$ has no meaning for $x\in Q_p$
and $\psi(x)\in C$. Also, there is no
possibility to define $p$-adic "momentum" or "Hamiltonian" operator.
In the real
case they are infinitesimal generators of space and time translations, but,
since $Q_p$ is disconnected field, these infinitesimal transformations
become meaningless.
However, finite transformations remain meaningful and the corresponding Weyl
and evolution operators are $p$-adically well defined.

The canonical commutation relation in the $p$-adic case can be represented by
the Weyl operators ($h=1$)
\begin{equation}
\hat Q_p(\alpha) \psi_p(x)=\chi_p(\alpha x)\psi_p(x),
\label{3.2}
\end{equation}
\begin{equation}
\hat K_p(\beta)\psi(x)=\psi_p(x+\beta) .
\label{3.3}
\end{equation}
Now, instead of the relation (\ref{3.1}) in the real case, we have
\begin{equation}
\hat Q_p(\alpha)\hat K_p(\beta)=\chi_p(\alpha\beta)
\hat K_p(\beta)\hat Q_p(\alpha)
\label{3.4}
\end{equation}
in the $p$-adic one.
It is possible to introduce the product of unitary operators
\begin{equation}
\hat W_p(z)=\chi_p(-\frac 1 2 qk)\hat K_p(\beta)\hat Q_p(\alpha), \quad
z\in Q_p\times Q_p,
\label{3.5}
\end{equation}
that is a unitary representation of the Heisenberg-Weyl group.
Recall that this group consists of the elements $(z,\alpha)$ with the
group product
\begin{equation}
(z,\alpha)\cdot
(z',\alpha ')=(z+z',\alpha+\alpha '+\frac{1}{2}
B(z,z')),
\label{3.6}
\end{equation}
where $B(z,z') = -kq'+qk'$ is a skew-symmetric bilinear
form on the phase space.
Dynamics of a $p$-adic quantum model is described by a unitary
evolution operator $U(t)$ without using the Hamiltonian operator. Instead of
that, the evolution operator has been formulated in terms of its kernel
${\cal K}_t(x,y)$
\begin{equation}
\label{3.7}
U_p(t)\psi(x)=\int_{Q_p}{\cal K}_t(x,y)\psi(y) dy.
\end{equation}
In this way,\cite{vol4} $p$-adic quantum mechanics is given by a triple
\begin{equation}
(L_2(Q_p), W_p(z_p), U_p(t_p)).
\label{3.8}
\end{equation}
Keeping in mind that standard quantum mechanics can be also given as the
corresponding triple, ordinary and $p$-adic
quantum mechanics can be unified in
the form of adelic quantum mechanics\cite{dra1,dra2}
\begin{equation}
(L_2({\cal A}), W(z), U(t)).
\label{3.9}
\end{equation}
$L_{2}({\cal A})$
is the Hilbert space on ${\cal A}$, $W(z)$ is a unitary representation of
the Heisenberg-Weyl group on $L_2({\cal A})$ and $U(t)$ is
a unitary representation of the
evolution operator on $L_2({\cal A})$.
The evolution operator $U(t)$ is defined by
\begin{equation}
U(t)\psi(x)=\int_{{\cal A}} {\cal K}_t(x,y)\psi(y)dy=\prod\limits_{v}{}
\int_{Q_{v}}{\cal K}_{t}^{(v)}(x_{v},y_{v})\psi^{(v)}(y_v) dy_{v}.
\label{3.10}
\end{equation}
The eigenvalue problem for $U(t)$ reads
\begin{equation}
U(t)\psi _{\alpha \beta} (x)=\chi (E_{\alpha} t)
\psi _{\alpha \beta} (x),
\label{3.11}
\end{equation}
where $\psi_{\alpha \beta}$ are adelic eigenfunctions,
$E_{\alpha }=(E_{\infty}, E_{2},..., E_{p},...)$ is the
corresponding adelic energy,
indices $\alpha$ and $\beta$ denote  energy levels and their
degeneration.  Any adelic eigenfunction has the form
\begin{equation}
\label{3.12}
\Psi_S(x) = \Psi_\infty(x_\infty)\prod_{p\in S}\Psi_p(x_p)
\prod_{p\not\in S}\Omega(| x_p|_p) , \quad x\in {\cal A},
\end{equation}
where $\Psi_{\infty}\in L_2(R)$,
$\Psi_{p}\in L_2(Q_p)$ are ordinary and $p$-adic eigenfunctions,
respectively. The $\Omega$-function is defined by (\ref{2.14}),
it is an element of the Hilbert space $L_2 (Q_p)$, and provides
convergence of the infinite product (\ref{3.12}). Note that
(\ref{3.12}) has the same form as (\ref{2.13}), but here all factors
are elements of the Hilbert spaces on $R$ and $Q_p$ of the same
quantum system. In an adelic eigenstate, $p$-adic eigenstates
are $\Omega(|x_p|_p)$ for all but a finite set $S$ of primes $p$.
Hence, the existence of $\Omega(|x_p|_p)$ for all or almost all
$p$ is a necessary condition for a quantum model to be adelic one.

For a fixed $S$, function $\Psi_S (x)$ in (\ref{3.12}) may be
regarded as an element of the Hilbert space $L_2({\cal A}(S))$,
where ${\cal A}(S)$ is a subset of adeles ${\cal A}$ defined in
Sec. 2. Moreover, $\Psi_\infty(x_\infty)$ and $\Psi_p(x_p)
\ (p\in S)$ may be not only eigenstates but also any element
of $L_2(R)$ and $L_2(Q_p)$, respectively. Then superposition
$\Psi (x)= \sum_S C(S) \Psi_S (x)$, where $\sum_S |C(S)|_\infty^2
=1$ and $\Psi_S \in L_2 ({\cal A}(S))$, is an element of
$L_2 ({\cal A})$.

A suitable way to calculate $p$-adic propagator ${\cal K}_p
(x'',t'';x',t')$ is to use Feynman's path integral method, i.e.
\begin{equation}
{\cal K}_p(x'',t'';x',t') = \int_{x',t'}^{x'',t''} \chi_p
\left( -\frac{1}{h} \int_{t'}^{t''} L(\dot{q},q,t) dt  \right) {\cal D}q.
\label{3.13}
\end{equation}
For quadratic Lagrangians it has been evaluated\cite{dra7,dra8}   in
the same way for real and $p$-adic cases, and the following
exact general expression is obtained:
\begin{equation}
\label{3.14}
{\cal K}_v(x'',t'';x',t')= \lambda_v \left( -
\frac{1}{2h}
\frac{\partial^2{\bar S}}{\partial x''\partial x'}
\right)
\left| \frac{1}{h}\frac{\partial^2{\bar S}}{\partial x''\partial x'}
\right|_v^{\frac{1}{2}}
\chi_v(-\frac{1}{h} {\bar S} (x'',t'';x',t')),
\end{equation}
where $\lambda_v$ functions satisfy properties (\ref{2.19}) and (\ref{2.20}),
and ${\bar S}(x'',t'';x',t')$ is the classical action.
When one has a system with more then one dimension with uncoupled spatial
coordinates, the total propagator is the product of the corresponding
one-dimensional propagators.
As an illustration of $p$-adic and adelic quantum-mechanical models,
the following one-dimensional systems with the
quadratic Lagrangians were considered: A free particle and a harmonic
oscillator,\cite{vol1,dra1,dra2} a particle in a constant field,\cite{dra9}
a free relativistic particle\cite{dra6}
and a harmonic oscillator with time-dependent frequency.\cite{dra10}

Adelic quantum mechanics takes into account ordinary as well as $p$-adic
quantum effects and may be regarded as a starting point for the construction of
a more complete superstring and M-theory. In the low-energy limit adelic
quantum mechanics becomes the ordinary one.\cite{dra6}

\section{$p$-Adic and Adelic Quantum Cosmology}
\noindent
Any  real space-time manifold in standard quantum cosmology
contains rational points which are dense in the field of real numbers.
These rational points, or some of them, may be completed with respect
to a distance induced by $p$-adic norm on $Q$ and one obtains $p$-adic
counterpart of this real manifold. Since this can be done for
every $p$, in this way, we
get an infinite ensemble of (real and $p$-adic) manifolds, which in the
form of the direct product usually  make  an adelic space-time. This
adelic space-time provides an arena for a simultaneous exhibition  of
real and $p$-adic aspects of gravitational and matter fields of the
same quantum cosmological model. According to the motivations (i)-(vi)
stated in Introduction, it is quite  reasonable to consider
our very early universe as an adelic quantum system.

Adelic quantum cosmology is an application of adelic quantum
theory to the universe as a whole.\cite{dra4} Adelic quantum theory unifies
both $p$-adic  and standard quantum theory.\cite{dra1} In the path
integral approach to standard quantum cosmology, the starting point
is Feynman's path integral method, i.e. the amplitude to go from one
state with intrinsic metric $h_{ij}'$ and matter configuration
$\phi'$ on an initial hypersurface $\Sigma'$ to
another state with metric $h_{ij}''$ and matter
configuration $\phi''$ on a final hypersurface
$\Sigma''$ is given by a functional integral
\begin{equation}
\langle h_{ij}'',\phi'',\Sigma''|
h_{ij}',\phi',\Sigma'\rangle_\infty =
\int {\cal D}{(g_{\mu\nu})}_\infty {\cal D}(\Phi)_\infty
\chi_\infty(-S_\infty[g_{\mu\nu},\Phi]),
\label {4.1}
\end{equation}
over all four-geometries $g_{\mu\nu}$ and matter configurations
$\Phi$, which interpolate between the initial and final
configurations.\cite{wil1} In this expression $S[g_{\mu\nu},\Phi]$ is
an Einstein-Hilbert action for the gravitational and matter
fields. This action can be calculated if we use metric in
the standard 3+1 decomposition
\begin{equation}
ds^2=g_{\mu\nu}dx^\mu dx^\nu=-(N^2 -N_i N^i)dt^2 + 2N_i dx^i dt
+ h_{ij} dx^i dx^j,
\label{4.2}
\end{equation}
where $N$ and $N_i$ are the lapse and shift functions.

To perform $p$-adic and adelic generalization we first make $p$-adic
counterpart of the action using form-invariance under change
of real to the $p$-adic number fields.
Then we generalize (\ref{4.1})
and introduce $p$-adic complex-valued cosmological amplitude
\begin{equation}
\langle h_{ij}'',\phi'',\Sigma''|
h_{ij}',\phi',\Sigma'\rangle_p =
\int{\cal D}{(g_{\mu\nu})}_p{\cal D}(\Phi)_p
\chi_p(-S_p[g_{\mu\nu},\Phi]),
\label {4.3}
\end{equation}
where $g_{\mu\nu}(x)$ and $\Phi (x)$ are the corresponding
$p$-adic counterparts of metric  and matter fields continually connecting
their values on $\Sigma'$ and $\Sigma''$. In its general aspects
$p$-adic functional integral (\ref{4.3}) mimics the usual Feynman
path integral (for one-dimensional case, see Refs. 2 and 21).
The definite integral in the classical action is understood as
the usual difference of the indefinite one
(without pseudoconstants) at final and initial points. The measures
${\cal D}(g_{\mu\nu})_p$ and ${\cal D}(\Phi)_p$ are related to
the real-valued Haar measure on $p$-adic spaces, and  the path integral
is the limit of a $k$-multiple integral when $k\to \infty$. There is no
natural ordering on $Q_p$, but one can define an appropriate
linear order.\cite{vol1}

Note that in (\ref{4.1}) and (\ref{4.3}) one has to take also a sum
over manifolds which have $\Sigma''$ and $\Sigma'$ as their boundaries.
Since the problem of topological classification of four-manifolds is
algorithmically unsovable it was proposed\cite{dra3} that summation should
be taken over algebraic manifolds. Our adelic approach supports this proposal,
since algebraic manifolds maintain all rational points under the interchange
of number fields $R$ and $Q_p$.

Since the space of all three-metrics and matter field configurations on a
three-surface, called superspace,  has infinitely many dimensions, one takes an
approximation. A useful approximation is to truncate the infinite degrees
of freedom to a finite number $q_\alpha(t)$, ($\alpha=1,2,...,n$). In this way,
one obtains a particular minisuperspace model. Usually, one
restricts the four-metric to be of the form (\ref{4.2}), with
$N^i=0$ and $h_{ij}$ as  functions  $q_\alpha(t)$.
For the homogeneous and isotropic cosmologies, the usual metric is a
Robertson-Walker one, of which the spatial sector has the form
\begin{equation}
h_{ij}dx^idx^j=a^2(t)d\Omega_3^2 = a^2(t) \left[
d\chi^2+\sin^2\chi(d\theta^2+\sin^2\theta d\varphi^2) \right].
\label {4.4}
\end{equation}
If we use also a single scalar field $\phi$, as a matter content
of the model, minisuperspace coordinates are $\{a,\phi\}$. More
generally, models can be homogeneous but also anisotropic ones,
and they will be here also considered. All such models can be
classified as: (i) Kantowski-Sachs models with spatial
topology $S^1\times S^2$ and
\begin{equation}
h_{ij}dx^idx^j=a^2(t)dr^2+ b^2(t)d\Omega_2^2,
\label {4.5}
\end{equation}
where $d\Omega_2^2$ is the metric  on the two-sphere,  and minisuperspace
coordinates are $\{a,b, \phi\}$; $(ii)$ Bianchi models, which are the most
general homogeneous cosmological models with a three-dimensional group
of isometries.
The three-metric of each of these models can be written  in the form
$ h_{ij}dx^idx^j=h_{ij}(t)\omega^i\otimes\omega^j,$
where $\omega^i$ are the invariant one-forms associated with the isometry
group. The simplest example is the Bianchi I model with
$\omega^1=dx,\ \omega^2=dy$ and $\omega^3=dz$, and
\begin{equation}
h_{ij}dx^idx^j=a^2(t)dx^2+b^2(t)dy^2+c^2(t)dz^2,
\label {4.6}
\end{equation}
where minisuperspace coordinates are $\{a,b,c,\phi\}$. For the
minisuperspace models, functional integrals in (\ref{4.1}) and
(\ref{4.3}) are reduced to functional integrals over three-metric and
configuration of matter fields, and to another usual integral over
the lapse function $N$. For the boundary condition
$q_\alpha(t'')=q_\alpha''$, $q_\alpha(t')=q_\alpha'$ in the gauge
$\dot N=0$, we have the $v$-adic minisuperspace propagator
\begin{equation}
\label{4.7}
\langle q_{\alpha}''|q_{\alpha}'\rangle_v
=\int_{G_v} dN {\cal K}_v(q_{\alpha}'',N;q_{\alpha}',0),
\end{equation}
where  $G_v$ is specified according to the adelic approach, i.e.
$G_\infty = R$ and $G_p = Z_p$ for all or almost all $p$, and
\begin{equation}
\label{4.8}
{\cal K}_v(q_\alpha'',N;q_\alpha',0)
=\int {\cal D}q_\alpha\chi_v(-S_v[q_\alpha]),
\end{equation}
is an ordinary quantum-mechanical propagator between fixed
minisuperspace coordinates ($q_\alpha',q_\alpha''$)  in a fixed 'time' $N$.
$S_v$ is the $v$-adic action of the minisuperspace
model, i.e.
\begin{equation}
\label{4.9}
S_v[q_\alpha]= \int_0^1dt N
\left[ \frac{1}{2N^2}
f_{\alpha\beta}(q)\dot q^\alpha\dot q^\beta-U(q) \right],
\end{equation}
where $f_{\alpha\beta}$ is a minisuperspace metric
$(ds^2_m=f_{\alpha\beta}dq^\alpha dq^\beta)$ with an indefinite
signature ($-,+,+,\dots$). This metric includes spatial
(gravitational) components and also matter variables for the
given model. It is worth emphasizing that in the adelic approach
the lapse function $N$
and minisuperspace coordinates $q_\alpha$ have adelic structure.
Also,  constants
and parameters must be the same rational numbers in $R$ and all $Q_p$.

The standard minisuperspace ground state wave
function in the Hartle-Hawking (no-boundary) proposal,\cite{har1}
will be attained if one performs a functional integration in the
Euclidean version of
\begin{equation}
\label{4.10}
\Psi_\infty[h_{ij}]= \int{\cal D}(g_{\mu\nu})_\infty
{\cal D}(\Phi)_\infty
\chi_\infty(-S_\infty[g_{\mu\nu},\Phi]),
\end{equation}
over all compact four-geometries $g_{\mu\nu}$ which induce $h_{ij}$ at the
compact three-manifold. This three-manifold is the only boundary of
all the four-manifolds.
If we generalize the Hartle-Hawking proposal to the $p$-adic minisuperspace,
then an adelic  Hartle-Hawking wave function is an infinite product
\begin{equation}
\label{4.11}
\Psi[h_{ij}]=
\prod_v\int{\cal D}(g_{\mu\nu})_v{\cal D}(\Phi)_v
\chi_v(-S_v[g_{\mu\nu},\Phi]),
\end{equation}
where the path integration must be performed over both, archimedean and
nonarchimedean geometries. If after evaluation of the corresponding
functional integrals we obtain as a result  $\Psi[h_{ij}]$ in the
form (\ref{3.12}), we will say that such cosmological model is an
adelic one.

As we shall see, a more successful $p$-adic generalization of the
minisuperspace cosmological
models can be  performed in the framework of $p$-adic and adelic
quantum mechanics\cite{dra5} without using the Hartle-Hawking proposal.
In such cases, we examine the conditions
under which some eigenstates of the evolution operator
(\ref{3.11}) exist.

\section{$p$-Adic Models in the Hartle-Hawking Proposal}
\noindent
The Hartle-Hawking proposal for the wave function of the universe is
generalized to $p$-adic case in Refs. 15 and 25. In this approach,
$p$-adic wave function is given by the integral
\begin{equation}
\label{5.1}
\Psi_p(q^\alpha)=\int_{G_p}dN{\cal K}_p(q^\alpha,N;0,0),
\end{equation}
where, according to the adelic structure of $N$, $G_p = Z_p$
(i.e. $|N|_p \leq 1$) for every or almost every $p$.

\subsection{Models of the de Sitter type}
\noindent
Models of the de Sitter type are models with
cosmological constant $\Lambda$ and without matter fields. We
consider two minisuperspace models of this type, with $D=4$ and
$D=3$ space-time dimensions. The corresponding real Einstein-Hilbert action
is\cite{hal1}
\begin{equation}
\label{5.2}
S= \frac{1}{16\pi G}\int_M d^Dx{\sqrt {-g}}(R-2\Lambda) +
\frac{1}{8\pi G}\int_{\partial M}d^{D-1}x\sqrt h K,
\end{equation}
where  $R$ is the scalar curvature of $D$-dimensional  manifold $M$,
$\Lambda$ is the cosmological constant, and $\ K$ is the trace of
the extrinsic curvature $K_{ij}$ on the boundary $\partial M$.
The metric for this model\cite{hal1} is of the
Robertson-Walker type
\begin{equation}
\label{5.3}
ds^2=\sigma^{D-2} [-N^2dt^2+a^2(t)d\Omega^2_{D-1}].
\end{equation}
In this expression $d\Omega^2_{D-1}$ denotes the metric on the unit
$(D-1)$-sphere, $\ \sigma^{D-2}={8\pi G}/$ $[{V^{D-1}(D-1)(D-2)}]$, where
$\ V^{D-1}$ is the volume of the unit $(D-1)$-sphere.

\subsubsection{The de Sitter model in $D=3$ dimensions}
\noindent
In the real $D=3$ case, the model is related to the multiple-sphere
configuration and  wormhole solutions.\cite{hal1}
$v$-adic classical action for this  model is
\begin{equation}
\label{5.4}
\bar S_v(a'',N; a',0)=
\frac{1}{2\sqrt\lambda} \left[ N\sqrt\lambda +
\lambda \left( \frac{2a''a'}{\sinh(N\sqrt\lambda)} -
\frac{a'^2+a''^2}{\tanh(N\sqrt\lambda)}\right)\right].
\end{equation}
Let us note that $\lambda$,  ($\lambda =\Lambda G^2$), denotes the rescaled
cosmological constant $\Lambda$.
Using (\ref{3.14}) for the propagator of this model we have
\begin{equation}
\label{5.5}
{\cal K}_v(a'',N;a',0) = \lambda_v
\left(-\frac{2\sqrt\lambda}{\sinh (N\sqrt\lambda)} \right)
\left| \frac{\sqrt\lambda}{\sinh(N\sqrt\lambda)} \right|_v^{1/2}
\chi_v(-\bar S_v(a'',N;a',0)).
\end{equation}
The $p$-adic Hartle-Hawking wave function is
\begin{equation}
\label{5.6}
\Psi_p(a)= \int_{|N|_p\leq 1} dN\frac{\lambda_p(-2N)}{|N|_p^{1/2}}
\chi_p \left( -\frac{N}{2}+\frac{\sqrt\lambda\coth(N\sqrt\lambda)}{2}a^2 \right),
\end{equation}
which after $p$-adic integration becomes
\begin{equation}
\label{5.7}
\Psi_p (a) =
\left\{\begin{array}{ll}
\Omega(|a|_p), & |\lambda|_p\leq p^{-2}, \quad p\neq 2,\\
\frac{1}{2}\Omega(|a|_2), & |\lambda|_2\leq 2^{-4}, \quad p=2.
\end{array}
\right.
\end{equation}

\subsubsection{The de Sitter model in $D=4$ dimensions}
\noindent
The de Sitter model in $D=4$ space-time dimensions may be described by the
metric\cite{hal2}
\begin{equation}
\label{5.8}
ds^2= \sigma^2 \left( -\frac{N^2}{q(t)}dt^2+q(t)d\Omega_3^2 \right),
\quad \sigma^2 =\frac{2G}{3\pi},
\end{equation}
and the corresponding action $S_v[q] = \frac{1}{2}\int_{t'}^{t''}dt N
\big(-\frac{\dot{q}^2}{4N^2} -\lambda q +1   \big)$, where
$\lambda = 2\Lambda G/(9\pi)$.
For $N=1$, the equation of motion  $\ddot{q} = 2\lambda$ has solution
$q(t)= \lambda t^2 + \big(\frac{q'' -q'}{T}-\lambda T \big)t +q'$,
where $q'' =q(t''), \ q' =q(t')$ and $T=t'' -t'$. Note that this
classical solution resembles motion of a particle in a constant field
and defines an algebraic manifold.
The choice of metric in the form (\ref{5.8})  yields quadratic
$v$-adic  classical action
\begin{equation}
\label{5.9}
\bar S_v(q'',T;q',0) =
\frac{\lambda^2T^3}{24} - [\lambda(q'+q'')-2 ]\frac{T}{4} -
\frac{(q''-q')^2}{8T}.
\end{equation}
 According to (\ref{3.14}),
 the corresponding propagator is
\begin{equation}\label{5.10}
{\cal K}_v(q'',T|q',0)=
\frac{\lambda_v(-8T)}{|4T|_v^{1/2}}
\chi_v(-\bar S_v(q'',T|q',0)).
\end{equation}
We obtain the $p$-adic Hartle-Hawking wave function  by the integral
\begin{equation}
\label{5.11}
\Psi_p(q)=
\int_{|T|_p\leq 1}
dT\frac{\lambda_p(-8T)}{|4T|_p^{1/2}}
\chi_p \left(
-\frac{\lambda^2T^3}{24}
+(\lambda q-2)\frac{T}{4}
+\frac{q^2}{8T} \right),
\end{equation}
and as a result we get\cite{dra4,dra11} also $\Omega(|q|_p)$ function with
the condition $\lambda=4\cdot 3\cdot l,\ l\in Z_p$.

The above $\Omega$-functions allow adelic wave functions of the form
(\ref{3.12}) for both $D=3$ and $D=4$ cases.
Since $|\lambda|_p \leq p^{-2}$ in (\ref{5.7}) for all $p\not=2$, it means that
$\lambda$ cannot be a rational number and consequently the above the de
Sitter minisuperspace model in $D=3$ space-time dimensions is not adelic one.
However $D=4$ case is adelic, because $\lambda = 4\cdot 3\cdot l$ is a rational
number when $l\in Z\subset Z_p$.

\subsection{Model with a homogeneous scalar field}
\noindent
To deal with the models of the de Sitter type  is very
instructive. Although these models are without matter content, they
are  in quantum cosmology of such significance as the model of
harmonic oscillator in quantum mechanics. However, it is also important
to consider models with some matter content.
In order to have a quadratic classical action, we use metric in the form\cite{hal3}
\begin{equation}
\label{5.12}
ds^2= \sigma^2\left(-N^2(t)\frac{dt^2}{a^2(t)}+a^2(t)d\Omega^2_3\right),
\end{equation}
the gravitational part of the action in the form (\ref{5.2}) (with $D=4$),
and the corresponding action for a scalar field as
\begin{equation}
\label{5.13}
S_{matter}=
-\frac{1}{2}\int_M d^4x\sqrt{-g}
\left[
g^{\mu\nu}\partial_\mu\Phi\partial_\nu\Phi+V(\Phi)
\right].
\end{equation}
After substitutions: $\Phi = \sqrt{3/(4\pi G)} \phi$, $V(\phi) =
\alpha \cosh (2\phi) + \beta \sinh (2\phi)$ and $x= a^2 \cosh (2\phi)$,
$y= a^2 \sinh (2\phi)$, we get  the classical action and propagator
$$
\bar S_p(x'',y'',N|x',y',0)=
\frac{\alpha^2-\beta^2}{24}N^3+
\frac{1}{4}(2-\alpha(x'+x'')-\beta(y'+y''))N
$$
\begin{equation}
\label{5.14}
+\frac{(y''-y')^2-(x''-x')^2}{8N},
\end{equation}
\begin{equation}
\label{5.15}
{\cal K}_p(x'',y'',N|x',y',0)
=\frac{1}{|4N|_p}
\chi_p(-\bar S_p(x'',y'',N;x',y',0)).
\end{equation}
As we have shown\cite{dra5} for this model, a $p$-adic Hartle-Hawking
wave function in the form of $\Omega$ - function  does not exist.
This leads to the conclusion that either the above
model is not adelic, or that $p$-adic generalization of the Hartle-Hawking
proposal is not an adequate one. However, if in the action (\ref{5.14})
we take $\beta=0,\ y=0$, then we get classical action for the de Sitter model
(\ref{5.9}), and such  model, as we showed it, is the adelic one.
The similar conclusion holds also for some other models in which
minisuperspace is not one-dimensional.
This is a reason to regard $p$-adic and adelic minisuperspace
quantum cosmology just as the correspondig application of $p$-adic and adelic
quantum mechanics without the Hartle-Hawking proposal.

\section{Minisuperspace Models in $p$-Adic and Adelic Quantum Mechanics}
\noindent
In this approach we investigate conditions under which
quantum-mechanical $p$-adic ground state exists in the form of
$\Omega$-function and some other typical eigenfunctions. This leads to
the desired result and it enables adelization of
many exactly soluble minisuperspace cosmological models, usually
with some restrictions on the parameters of the models.

The necessary
condition for the existence of an adelic quantum model is the existence of
$p$-adic  ground state $\Omega(|q_\alpha|_p)$  defined by (\ref{2.14}),
i.e.
\begin{equation}
\label{6.1}
\int_{|{q_\alpha}'|_p\leq1}{\cal K}_p
({q_\alpha}'',N;{q_\alpha}',0)d{q_\alpha}'=
\Omega(|{q_\alpha}''|_p).
\end{equation}
Analogously, if a system  is in
the state
$\Omega(p^\nu|q_\alpha|_p)$, where $\Omega(p^\nu|q_\alpha|_p) = 1$ if
$|q_\alpha|_p \leq p^{-\nu}$ and $\Omega(p^\nu|q_\alpha|_p) = 0$ if
$|q_\alpha|_p > p^{-\nu}$ , then its kernel must satisfy equation
\begin{equation}
\label{6.2}
\int_{|{q_\alpha}'|_p\leq p^{-\nu}}{\cal K}_p
({q_\alpha}'',N;{q_\alpha}',0)d{q_\alpha}'=
\Omega(p^\nu|q_\alpha''|_p).
\end{equation}
If $p$-adic  ground state is of the form of the
$\delta$-function, where  $\delta$-function is defined as
$\delta (a-b) = 1$ if $a=b$ and $\delta (a-b) =0$ if $a\neq b$,
then  the corresponding kernel of the model has to satisfy equation
\begin{equation}
\label{6.3}
\int_{|q'_\alpha|_p =p^\nu}{\cal K}_p(q_\alpha'',T;q',0)dq_\alpha'=
\delta(p^\nu-|q_\alpha''|_p).
\end{equation}
Equations (\ref{6.1}) and (\ref{6.2}) are usual $p$-adic vacuum
ingredients of the adelic eigenvalue problem (\ref{3.11}),
i.e. $\int_{Q_p} {\cal K}_p (x'',t;x',0)
\Psi_p(x') dx' = \chi_p (Et)\Psi_p(x'')$, where $\chi_p(Et)=1$ in
the vacuum state $\{ Et\}_p = 0$. The above $\Omega$ and $\delta$
functions do not make a complete set of $p$-adic eigenfunctions, but
they are very simple and illustrative. Since these functions have
finite supports, the ranges of integration in (\ref{6.1})-(\ref{6.3})
are also finite. The lapse function $N$ is under the kernel
${\cal K}(q_\alpha'',N;q_\alpha',0)$ and is restricted to some values
on which eigenfunctions do not depend explicitely.

In the following, we apply (\ref{6.1})-(\ref{6.3})
to some minisuperspace models.

\subsection{Models of the de Sitter type}
\subsubsection{The de Sitter model in $D=3$ dimensions}
\noindent
By  application of the above exposed formalism of $p$-adic quantum mechanics
in the form (6.1) and (6.2), for this model we found\cite{dra5}
the ground state
\begin{equation}
\label {6.4}
\Psi_p(a) =
\left\{\begin{array}{lr}
\Omega(|a|_p), & |N|_p\leq 1,\quad |\lambda|_p\leq \frac{1}{p^2}, \quad  p\neq 2, \\
\Omega(|a|_2), & |N|_2\leq\frac{1}{4},\quad |\lambda|_2\leq 4, \quad p=2,
\end{array}
\right.
\end{equation}
 and  also
\begin{equation}
\label{6.5}
\Psi_p(a)=
\left\{\begin{array}{lr}
\Omega(p^\nu|a|_p), & |N|_p\leq p^{-2\nu},\quad |\lambda|_p\leq p^{4\nu -2},
\quad p\neq 2,\\
\Omega(2^\nu|a|_2), & |N|_2\leq 2^{-2-2\nu},\quad |\lambda|_2\leq 2^{4\nu +2},
\quad p=2,
\end{array}
\right.
\end{equation}
where $\nu=1,2,3,\cdots$\ . For simplicity, in the sequel the upper and lower
row will be related to the $p\neq 2$ and $p=2$ cases, respectively.

The existence of the ground state in the
form of the $\delta$-function may be investigated by the Eq.
(\ref{6.3}), i.e.
\begin{equation}
\label{6.6}
\int_{Q_p}{\cal K}_p(a'',N;a',0)\delta(p^\nu-|a'|)da'=
\delta(p^\nu-|a''|),
\end{equation}
with the kernel (\ref{5.5}), that leads to the equation
$$
\lambda_p \left(
-\frac{\sqrt\lambda}{2\sinh(N\sqrt\lambda)}
\right) \left|
\frac{\sqrt\lambda}{\sinh(N\sqrt\lambda)}
\right|_p^{1/2}\chi_p
\left(
-\frac{N}{2}+\frac{\sqrt\lambda}{2\tanh(N\sqrt\lambda)}{a''}^2
\right)
$$
\begin{equation}
\label{6.7}
\times \int_{|a'|_p=p^\nu}\chi_p
\left(
\frac{\sqrt\lambda}{2\tanh(N\sqrt\lambda)}{a'}^2
-\frac{\sqrt\lambda}{\sinh(N\sqrt\lambda)}{a''}a'
\right)da'
=\delta(p^\nu- |a''|_p).
\end{equation}
The above integration is performed over $p$-adic sphere
 with the radius $p^\nu$ and for
$|\frac{N}{2}|_p\le p^{2\nu-2} $, $\nu=1,0,-1,-2,\dots$\ . As a result, on
the left hand side we have
$$ \chi_p \left(
-\frac{N}{2}+\frac{\sqrt\lambda}{2}\tanh(N\sqrt\lambda){a''}^2
\right). $$
To have an equality, the argument of the additive character must be
equal or less than unity. This requirement leads to the condition $$ \left|
\frac{\sqrt\lambda\tanh(N\sqrt\lambda){a''}^2}{2}
\right|_p\leq p^{4\nu-2}|\lambda|_p\leq 1,\enskip
\enskip|\lambda|_p\leq p^{2-4\nu}. $$
This  (for the
$p$-adic norms of $N$ and $\lambda$) is also related to the
domain of convergence of the analytic function $\tanh x$, $$
|N\sqrt\lambda|_p = |N|_p|\lambda|_p^{1/2}\leq p^{2\nu-2}\cdot
p^{1-2\nu} =p^{-1}. $$ If $p=2$, then condition
$|N|_2\leq 2^{2\nu-3}$ holds, for $\nu=1,0,-1,-2,\cdots$, and
we are in the domain of convergence. Finally, we also
conclude that $p$-adic ground state
\begin{equation}
\label{6.8}
\Psi_p(a)=
\left\{\begin{array}{rl}
\delta(p^\nu-|a|_p), & |N|_p\leq p^{2\nu-2},\quad|\lambda|_p\leq p^{2-4\nu},\\
\delta(2^\nu-|a|_2), & |N|_2\leq2^{2\nu-3},\quad|\lambda|_2\leq 2^{2-4\nu},
\end{array}
\right.
\end{equation}
exists for $\nu=1,0,-1,-2,\cdots$.

This de Sitter model is $p$-adic but it is not an adelic one for
the same reasons as in the above Hartle-Hawking approach.

\subsubsection{The de Sitter model in $D=4$ dimensions}
\noindent
Here we start using the Eqs. (\ref{5.8})-(\ref{5.10}).
As it was already shown,\cite{dra5} the ground states
for this model exist in the forms
\begin{equation}
\label{6.9}
\Psi_p(q)=
\left\{\begin{array}{rl}
\Omega(|q|_p), & |T|_p\leq1,\quad\lambda=4\cdot 3\cdot l,\quad l\in Z, \\
\Omega(|q|_2), & |T|_2\leq\frac{1}{2},\quad\lambda=4\cdot 3\cdot l,\quad l\in Z,
\end{array}
\right.
\end{equation}
\begin{equation}
\label{6.10}
\Psi_p(q)=
\left\{\begin{array}{rl}
\Omega(p^\nu|q|_p), & |T|_p\leq p^{-2\nu},\quad |\lambda|_p\leq |3|_p^{1/2}p^{3\nu},
\quad \nu= 1,2,3,\cdots, \\
\Omega(2^\nu|q|_2), & |T|_2\leq 2^{-2\nu},\quad |\lambda|_2\leq 2^{3\nu-1},
\quad \nu=1,2,3\cdots.
\end{array}
\right.
\end{equation}

Looking for the existence of the $p$-adic ground state in the
form of the  $\delta$-function, we have to solve the integral equation
$$
\frac{\lambda_p(-8T)}
{|4T|_p^{1/2}}
\chi_p
\left(
-\frac{\lambda^2T^3}{24}-\frac{T}{2}+\frac{\lambda q''T}{4}+
\frac{{q''}^2}{8T}
\right)
$$
\begin{equation}
\times
\int\limits_{|q'|_p=p^\nu}\chi_p
\left(
\frac{{q'}^2}{8T}+\left(
\frac{\lambda T}{4}-\frac{q''}{4T}
\right)q'
\right)dq'
=\delta(p^\nu-|q''|_p).
\label{6.11}
\end{equation}
After the corresponding integration, for
the left hand side of the previous equation, we obtain
$$ \chi_p \left(
-\frac{\lambda^2T^3}{6}-\frac{T}{2}+\frac{\lambda q''}{2}T \right).$$  
By the very similar analysis for the parameter $\lambda$, we
get $|\lambda|_p\leq p^{2-3\nu}$, and finally
\begin{equation}
\label{6.12}
\Psi_p(q)=
\left\{\begin{array}{rl}
\delta(p^\nu-|q|_p), & |T|_p\leq p^{2\nu-2},\quad|\lambda|_p\leq p^{2-3\nu},\\
\delta(2^\nu-|q|_2), & |T|_2\leq2^{2\nu-1},
\quad|\lambda|_2\leq2^{-3\nu},
\end{array}
\right.
\end{equation}
where   $\nu=1,0,-1,-2,\cdots$ if $p\neq 2$,
and  $\nu=0,-1,-2,\cdots$ if $p=2$.

This de Sitter model is adelic one. It allows eigenfunctions of the form
(\ref{3.12}), where for $p\in S$ states $\Psi_p(x_p)$ may be some solutions
(\ref{6.10}) and (\ref{6.12}) with appropriately chosen $l\in Z$ in
$\lambda = 3\cdot 4\cdot l$.

\subsection{Model with a homogeneous scalar field}
\noindent
This is an adelic two-dimensional minisuperspace model  with two
decoupled degrees of freedom. On the basis of (\ref{5.12})-(\ref{5.15})
and (\ref{6.1})-(\ref{6.2}), the ground state with $\Omega$-type functions is
\begin{equation}
\label{6.13}
\Psi_p(x,y)=
\left\{\begin{array}{rl}
\Omega(|x|_p)\ \Omega(|y|_p), & |N|_p\leq1, \quad \alpha=4\cdot3\cdot l_1, \quad
\beta=4\cdot3\cdot l_2,\\
\Omega(|x|_2)\ \Omega(|y|_2), & |N|_2\leq\frac{1}{2}, \quad
\alpha=4\cdot3\cdot l_1,\quad \beta=4\cdot3\cdot l_2,
\end{array}
\right.
\end{equation}
where $\ l_1,l_2\in Z$, and also
\begin{equation}
\label{6.14}
\Psi_p(x,y)=
\left\{\begin{array}{rl}
\Omega(p^\nu|x|_p)\ \Omega(p^\mu|y|_p),& |N|_p\leq p^{-2\nu},\quad |N|_p
\leq p^{-2\mu},  \\
\Omega(2^\nu|x|_2)\ \Omega(2^\mu|y|_2),& |N|_2\leq 2^{-2\nu}, \quad |N|_2
\leq 2^{-2\mu},
\end{array}
\right.
\end{equation}
with
$ |\alpha|_p\leq |3|_p^{1/2}p^{3\nu},\ |\beta|_p\leq |3|_p^{1/2}p^{3\mu}$
(if $p\neq 2$) and $ |\alpha|_2\leq 2^{3\nu -1},\ |\beta|_p\leq 2^{3\mu-1}$,
where $\nu ,\mu = 1,2,3,\cdots$.
As in the previous cases, we also investigate the existence of
the vacuum state of the form
$\delta(p^\nu-|x|_p)\ \delta(p^\nu-|y|_p).$
After the very similar calculations in Subsec. 6.1, we find
$p$-adic wave function for the ground state
\begin{equation}
\Psi_p(x,y) =
\label{6.15}
\left\{\begin{array}{rl}
\delta(p^\nu-|x|_p)\ \delta(p^\mu-|y|_p), &
     |N|_p\leq p^{2\nu-2},\quad |N|_p\leq p^{2\mu-2},\\
\delta(2^\nu-|x|_2)\ \delta(2^\mu-|y|_2), &
     |N|_2\leq2^{2\nu-1},\quad |N|_2\leq 2^{2\mu-1},
\end{array}
\right.
\end{equation}
with $|\alpha|_p\leq p^{2-3\nu}, \ |\beta|_p\leq p^{2-3\mu} $, and
$|\alpha|_2\leq 2^{-3\nu}, \ |\beta|_2\leq 2^{-3\mu}$,
where $\nu,\mu=0,-1,-2,\cdots$.

\subsection{Anisotropic Bianchi Model with three scale factors}
\noindent
In this adelic case we start with metric\cite{ish1}
\begin{equation}
\label{6.16}
ds^2=\sigma^2\left[-\frac{N^2(t)}{a^2(t)}dt^2
+ a^2(t)dx^2 +b^2(t)dy^2+c^2(t)dz^2\right].
\end{equation}
It leads to the action
\begin{equation}
\label{6.17}
S_v[a,b,c]=\frac{1}{2}\int^1_0dt
\left[-\frac{a}{N}(\dot a\dot bc +a\dot b\dot c+\dot ab\dot c)
-Nbc\lambda\right].
\end{equation}
By means of the substitution
\begin{equation}
\label{6.18}
x=\frac{bc+a^2}{2},\quad y=\frac{bc-a^2}{2},\quad\dot z^2=a^2\dot b\dot c
\end{equation}
we obtain the quadratic classical action and propagator in the form
$$
\bar S_v(x'',y'',z'',N;x',y',z',0)=
$$
\begin{equation}
-\frac{1}{4N}\left[(x''-x')^2-(y''-y')^2+2(z''-z')^2\right]
-\frac{\lambda N}{4}\left[(x'+x'')+(y'+y'')\right],
\label{6.19}
\end{equation}
\begin{equation}
\label{6.20}
{\cal K}_\upsilon(x'',y'',z'',N;x',y',z',0)=
\frac{\lambda_\upsilon(-2N)}{\left|4^{1/3}N\right|_\upsilon^{3/2}}
\chi_\upsilon\left(-\bar S_\upsilon(x'',y'',z'',N;x',y',z',0)
\right).
\end{equation}

By the above way, one gets the $p$-adic eigenstates
\begin{equation}
\label{6.21}
\Psi_p(x,y,z)=
\left\{\begin{array}{rl}
\Omega(|x|_p)\ \Omega(|y|_p)\ \Omega(|z|_p), & |N|_p\leq1,
\quad|\lambda|_p\leq1,\\
\Omega(|x|_2)\ \Omega(|y|_2)\ \Omega(|z|_2), & |N|_2\leq\frac{1}{4},
\quad |\lambda|_2\leq2,
\end{array}
\right.
\end{equation}
and
\begin{equation}
\label{6.22}
\Psi_p(x,y,z) =
\left\{\begin{array}{rl}
\Omega(p^{\nu_1}|x|_p)\ \Omega(p^{\nu_2}|y|_p)\ \Omega(p^{\nu_3}|z|_p), &
 \enskip \\
\Omega(2^{\nu_1}|x|_2)\ \Omega(2^{\nu_2}|y|_2)\ \Omega(2^{\nu_3}|z|_2),
&
\end{array}
\right.
\end{equation}
with conditions: $|N|_p\leq p^{-2\nu_1},\ |N|_p\leq p^{-2\nu_2},\
|N|_p\leq p^{-2\nu_3}, \quad |\lambda|_p\leq p^{3\nu_1},
\ |\lambda|_p\leq p^{3\nu_2}$,
(if $p\neq 2$) and  $|N|_2\leq2^{-2\nu_1-1}, \ |N|_2\leq2^{-2\nu_2-1},\ |N|_2\leq 2^{-2\nu_3-2},
\quad |\lambda|_2\leq 2^{3\nu_1+1}, \ |\lambda|_2\leq 2^{3\nu_2+1}$,
 where $\nu_i = 1,2,3,\cdots$.

For this model there also exist ground states
\begin{equation}
\label{6.23}
\Psi_p(x,y,z) =
\left\{\begin{array}{rl}
\delta(p^{\nu_1}-|x|_p)\ \delta(p^{\nu_2}-|y|_p)\ \delta(p^{\nu_3}-|z|_p), & \\
\delta(2^{\nu_1}-|x|_2)\ \delta(2^{\nu_2}-|y|_2)\ \delta(2^{\nu_3}-|z|_2), &
\end{array}
\right.
\end{equation}
with conditions: $  |N|_p\leq p^{2\nu_1-2},\ |N|_p\leq p^{2\nu_2-2},\
|N|_p\leq p^{2\nu_3-2}, \quad |\lambda|_p\leq p^{2-3\nu_1}$,
$|\lambda|_p\leq p^{2-3\nu_2}$,
and $ |N|_2\leq 2^{2\nu_1-3}, \ |N|_2\leq 2^{2\nu_2-3},\
|N|_2\leq 2^{2\nu_3-3},\quad   |\lambda|_2\leq2^{2-3\nu_1}$,
$|\lambda|_2\leq2^{2-3\nu_2}$,
where $\nu_1 ,\nu_2=1,0,-1,-2,\cdots,\ \ \nu_3 \in Z$.

\subsection{Some two dimensional models}
\noindent
There is a class of two-dimensional minisuperspace models
which, after some transformations,  obtain the form of two
oscillators.\cite{lou1,pag1}
These models are: the isotropic Friedmann model with conformally  and
minimally coupled scalar field, and the ani\-so\-tro\-pic vacuum
Kantowski-Sachs model. For all these three models  the corresponding
action may be written as
\begin{equation}
\label{6.24}
S= {1\over 2} \int_0^1 dtN \left[-{{\dot x^2}\over {N^2}}+
{{\dot y^2}\over {N^2}} + x^2-y^2\right],
\end{equation}
i.e. this is the action for two linear oscillators, but one of
them has negative energy. This classical action leads to the propagator
\begin{equation}
\label{6.25}
{\cal K}_p(y'',x'',N;y',x',0)= {1\over {|N|_p}}
\chi_p \left(\frac{{x''}^2+{x'}^2-{y'}^2-{y''}^2}
{2\tan N} + {{y'y''-x'x''}\over {\sin N}}\right).
\end{equation}

The linear harmonic oscillator was  analyzed  from
$p$-adic, as well as from the adelic point of
view.\cite{vol1,dra1,dra2} One can  show that  in the $p$-adic
region of convergence of analytic functions $\sin N$ and $\tan N$,
which is $G_p=\{N\in Q_p:|N|_p\leq|2p|_p\}$, exist vacuum
states $\Omega(|x|_p)$ $\Omega(|y|_p)$,
$\ \Omega(p^\nu|x|_p)$ $\Omega(p^\mu|y|_p), \ \nu,\mu =1,2,3,\cdots$, and also
\begin{equation}
\label{6.26}
\Psi_p(x,y) =
\left\{\begin{array}{rl}
\delta(p^\nu-|x|_p)\ \delta(p^\mu-|y|_p), & |N|_p\leq p^{2\nu-2},\quad
|N|_p\leq p^{2\mu-2}, \\
\delta(2^\nu-|x|_2)\ \delta(2^\mu-|y|_2), & |N|_2\leq 2^{2\nu-3}, \quad
|N|_2\leq 2^{2\mu-3},
\end{array}
\right.
\end{equation}
where $\nu,\mu= 0,-1,-2,\cdots$. This is another example of $p$-adic and
adelic minisuperspace cosmological model.

\section{Concluding Remarks}
\noindent
In this paper, we find  application of $p$-adic numbers in quantum
cosmology  very promising. It gives new and more complete insights
into the space-time structure at the Planck scale.

In the Hartle-Hawking approach the wave function of a spatially
closed universe is defined by Feynman's path integral method.
The action is a functional of the gravitational and matter fields, and
path integration is performed over all compact real four-metrics
connecting two three-space states. Accordingly, the present adelic
Hartle-Hawking proposal  extends the ordinary one to the
all corresponding compact $p$-adic metrics. Unfortunately, it does not lead
to the adequate adelic generalization for a wide class
of the minisuperspace models.

However, the consideration of minisuperspace models in
the framework of adelic quantum mechanics gives the appropriate
 adelic generalization. Moreover, we can conclude that all
the above adelic models lead to the new space-time picture in the vicinity
of the Planck length.

Namely, for all the above adelic models there exist adelic ground states
of the form
\begin{equation}
\label{7.1}
\Psi_S(q^1,...,q^n)=\prod_{\alpha=1}^n  \Psi_\infty(q^\alpha_\infty)
\prod_{p\in S} \Psi_p (q_p^\alpha) \prod_{p\not\in S} \Omega(|q^\alpha_p|_p),
\end{equation}
\noindent
where $\Psi_\infty(q^\alpha_\infty)$ are the corresponding real counterparts
of the wave functions of the universe, and $\Psi_p (q_p^\alpha)$ are proportional
to $\Omega(p^\nu |q_p^\alpha|_p)$ or $\delta (p^\nu -|q_p^\alpha|)$
eigenfunctions with the corresponding normalization factors.
Adopting the usual probability interpretation of the wave function (\ref{7.1}),
we have
\begin{equation}
\label{7.2}
\left|
\Psi_S(q^1,...,q^n)
\right|_\infty^2 =
\prod_{\alpha=1}^n \left| \Psi_\infty(q_\infty^\alpha) \right|_\infty^2
\prod_{p\in S} |\Psi_p(q_p^\alpha)|_\infty^2  \prod_{p\not\in S}
\Omega(|q_p^\alpha|_p),
\end{equation}
\noindent because $(\Omega(|q^\alpha|_p))^2=\Omega(|q^\alpha|_p)$.

As a consequence of $\Omega$-function properties, at the rational points
$q^1,\cdots,q^n$ and $S =\emptyset$, we have
\begin{equation}
\label{7.3}
\left|
\Psi(q^1,\dots,q^n)
\right|_\infty^2=
\cases
{\prod_{\alpha =1}^n\left|\Psi_\infty(q^\alpha)\right|_\infty^2, &$q^\alpha\in Z$,\cr
0, &$q^\alpha\in Q\backslash Z$.\cr}
\end{equation}
This result leads to some discretization of minisuperspace
coordinates $q^\alpha$, because for all rational points density probability
is nonzero only in the integer points of $q^\alpha$. Keeping in mind that
$\Omega$-function is invariant with respect to the Fourier transform, this
conclusion is also valid for the momentum space. Note that
this kind of discreteness depends on adelic quantum state of the
universe. When some $p$-adic states (for $p \in S$) are
different from $\Omega(|q^\alpha|_p)$,
then the above adelic discretennes becomes less transparent.

There is also some discreteness of the cosmological constant. Namely,
the parameter $\lambda$ is proportional to $\Lambda$, and  consequently
to the vacuum energy density. Since $\lambda $ may have only integer values
(see e.g. (\ref{6.9}) for the de Sitter model), it  follows that vacuum energy
of the universe belongs to a discrete spectrum, which depends on its
adelic quantum state.

It is worth noting that investigation of quantum properties of the de Sitter
space is an actual subject. In particular, it has been argued\cite{ban1}
and discussed\cite{wit1} that it is possible that the Hilbert space of a quantum
de Sitter space has a finite dimension.  According to our results,
the corresponding adelic Hilbert space is of the infinite dimension.
Discreteness\cite{aiv1} of the Planck constant in the de Sitter space  and
the dS/CFT correspondence\cite{str1} have been also investigated.

Performing the integration in (\ref{7.2}) over all the $p$-adic spaces, and having
in mind that eigenfunctions should be normed to unity, one recovers the standard
effective model over real space. However, if the region of integration
is over only some parts of $p$-adic spaces then the adelic approach
manifestly exhibits $p$-adic
quantum effects. Since the Planck length is here the natural one, the adelic
minisuperspace models refer to the Planck scale.

\nonumsection{Acknowledgements}
\noindent
G.S.Dj. is partially supported by DFG Project "Noncommutative space-time
structure - Cooperation with Balkan Countries". The work of B.D. and I.V.V.
was supported in part by RFFI grant 990100866. We would like to thank the
referee for suggestions which led to better presentation of some parts
of the paper.

\nonumsection{References}

\end{document}